\documentclass[review]{elsarticle}

\usepackage{hyperref}
\makeatletter
\def\ps@pprintTitle{%
 \let\@oddhead\@empty
 \let\@evenhead\@empty
 \def\@oddfoot{\centerline{\thepage}}%
 \let\@evenfoot\@oddfoot}
\makeatother

\begin{document}

\begin{frontmatter}

\title{The Cause of a Pulsar Glitch}

\author{Thando Nkomozake}
\address{Department of Mathematics and Applied Mathematics,\\
University of Cape Town,\\
Private Bag, Rondebosch, 7700, South Africa}


\begin{abstract}
\noindent In this paper we present a new mechanism for rationalizing, explaining and predicting pulsar glitches. Our new mechanism is based upon the concepts of type II superconductivity and the Bradlow bound.

\end{abstract}

\end{frontmatter}

\section{INTRODUCTION}

\noindent It was pointed out by Thomas Gold in 1968 that the rotational energy of a pulsar would decrease due to magnetic dipole radiation, resulting in the pulsar slowing down. The spin down is gradual and largely predictable. However, pulsars have a timing irregularity, where they experience a sudden and spectacular increase in rotation velocity, which is called a pulsar glitch. Up until the present day, there has been hundreds of pulsar glitches that have been detected, and their spin frequency $\Omega$ have relative increases that lie between $\Delta \Omega / \Omega \approx 10^{-11}$ and $\Delta \Omega / \Omega \approx 10^{-5}$. \\[0.1 in] 
After the first glitches were observed in the Vela pulsar, around 1969, it was said that a superfluid component in the interior of the pulsar was responsible and that a weak coupling exists between the normal component and the electromagnetic emission. It was suggested that such a system stores up angular momentum and then releases it to cause the glitch \cite{seveso2015advances}. This mechanism is called vortex pinning and was first suggested by P.W. Anderson and N. Itoh. It states that: in order for a superfluid to rotate, it forms a configuration of quantized vortices that carry the circulation of the superfluid. In the neutron star crust the vortices are strongly attracted and pinned to the nuclear lattice and cannot move outward. This behaves like a reservoir for angular momentum. Due to electromagnetic emission, the crust of the neutron star spins down and as a result of that there's a lag that develops between the normal component and the superfluid. This lag leads to a hydrodynamical force that is called the magnus force, which acts on the vortices. When this lag reaches a critical point, the pinning force is no longer able to contrast the magnus force. At this point the vortices unpin and they transfer their angular momentum to the crust and this causes the glitch. Although this theory is generally accepted, it still has some unanswered questions. It doesn't give us the location of the angular momentum reservoir in the star and doesn't tell us what percentage of the star it constitutes. It also doesn't tell us what is responsible for the angular momentum transfer and where in the star this coupling occurs. Several mechanisms have been considered in an attempt to answer these questions and all have been unsatisfactory. \\[0.1 in]
In this paper we present a different mechanism for the pulsar glitch. To the best of our knowledge our approach is new. The currently accepted view is that the outer-core of a neutron star is composed of a neutron superfluid with electrons, muons and type II superconducting protons immersed in the superfluid. In order to have a simple mechanism for pulsar glitching, we slightly alter that view. Instead of thinking about the superconducting protons as being immersed in the superfluid, we think of them as forming a solid type II superconducting interface, which we call ``the outer-core surface", and that lies between the inner-crust and the outer-core of the pulsar. In our opinion, type II superconductors are generally more interesting than type I superconductors. We hold this opinion because, a type I superconductor has only one critical magnetic field, at which it becomes normal conducting. On the other hand a type II superconductor doesn't just go from a superconducting to a normal conducting state at a single critical field. Instead it has a lower critical field and an upper critical field, which allows for the formation and penetration of magnetic vortices, see Figure 1. \\[0.1 in]
Therefore our approach allows us to apply the topological idea of the Bradlow bound and also relate it to the upper critical field of the outer-core surface. With our approach, the mechanism for pulsar glitching is based on the Bradlow bound of the outer-core surface which corresponds to its upper critical field. In this paper we present our mechanism in the form of three assertions or statements that are given as the headings of sections 2, 3 and 4. In each of these sections we present a number of results in bullet point form, in support of the statement made in the heading of that section. Most of the results we present in these sections are derived from various literature we reviewed and cited. We also add our own deductions in support of the mechanism we propose. \\[0.1 in]
Because of the way we structured this paper there is some overlap in notation: \begin{itemize}
    \item  $\lambda  \equiv \lambda_{p} $
    
    \item $\xi_{ft} \equiv \xi_{p}     $
    
    \item $m_{p}^{*} \equiv m_{p}  $
    
    \item $ \mathcal{N}_{ft} \equiv n_{f}  $
    
    \item $ d_{ft} \equiv l_{f} $
    
\end{itemize}

\noindent  We use: magnetic flux tube, proton vortex and magnetic vortex interchangeably. We also use: neutron star and pulsar interchangeably.

\section{THE SURFACE OF THE OUTER-CORE OF A PULSAR IS A TYPE II SUPERCONDUCTOR}

\begin{itemize}
\item The nucleons in the neutron stars' core form Cooper pairs and exhibit macroscopic quantum behavior \cite{graber2017fluxtube}. 
\item There are superconducting protons in the outer-core that show type II properties \cite{graber2017fluxtube}.

\item The magnetic field in the outer-core is no longer locked to the charged plasma but instead confined to magnetic flux tubes \cite{graber2017fluxtube}.
\item The type of superconductivity in the outer-core depends on the characteristic length scales involved. An estimate of the Ginzburg-Landau parameter for the superconducting protons in the outer-core is given by \cite{graber2017fluxtube},
\begin{equation} \kappa = \frac{\lambda}{\xi_{ft}} \approx 3.3 \bigg(   \frac{m_{p}^{*}}{m_{b}}  \bigg )^{3/2 } \rho_{14}^{-5/6} \bigg( \frac{x_{p}}{0.05}   \bigg)^{-5/6} \bigg( \frac{T_{cp}}{10^{9} K} \bigg),   \end{equation} where $\lambda$ is the penetration depth of the magnetic flux tubes, $\xi_{ft}$ is the coherence length of the superconducting protons, $m_{p}^{*}$ is the proton effective mass, $m_{b}$ is the baryon mass, $\rho_{14} = \rho / (10^{14} g \cdot cm^{-3})$ is the normalized total mass density, $x_{p}$ is the proton fraction, $T_{cp} \approx 10^{9}-10^{10} K$ is the proton transition temperature. Equation (1) gives a value that is larger than the critical value $\kappa_{crit} = 1/ \sqrt{2}$, and this suggests that the neutron star's interior is in a type II state.

\item The lower critical field for the superconducting protons is given by \cite{graber2017fluxtube, sinha2015upper}, \begin{equation} B_{c1} \approx      1.9 \times 10^{14}  \bigg(   \frac{m_{b}}{m_{p}^{*}}  \bigg ) \rho_{14} \bigg( \frac{x_{p}}{0.05}   \bigg) \ G,        \end{equation} and the upper critical field is given by \cite{graber2017fluxtube, buckley2004vortices}, \begin{equation} B_{c2} \approx
2.1 \times 10^{15} \bigg(   \frac{m_{p}^{*}}{m_{b}}  \bigg )^{2 } \rho_{14}^{-2/3} \bigg( \frac{x_{p}}{0.05}   \bigg)^{-2/3} \bigg( \frac{T_{cp}}{10^{9} K} \bigg)^{2} G.
\end{equation}

\item The outer-core is in a metastable type II state and penetrated by flux tubes \cite{graber2017fluxtube}.

\item The quantized flux tubes on the outer-core type II region are arranged in a hexagonal array. Each flux tube has a unit of flux, $\phi_{0} \approx 2 \times 10^{-7} G \cdot cm^{2}$. The macroscopic magnetic induction $B$ in the star's core is obtained by summing all individual flux quanta \cite{graber2017fluxtube}.

\item The flux tube surface density is given by \cite{graber2017fluxtube}, \begin{equation} \mathcal{N}_{ft} \approx 4.8 \times 10^{18} B_{12} \ cm^{-2},  \end{equation} and the inter-flux tube distance is given by \cite{graber2017fluxtube}, \begin{equation}    d_{ft} \approx 4.6 \times 10^{-10} B_{12}^{-1/2} \ cm,             \end{equation} 

\end{itemize} where $B_{12} = B / 10^{12}\ G$.\\

 \noindent Based on the points above, we now form a similar but slightly different point of view. Instead of thinking about the superconducting protons as floating or immersed in the neutron superfluid in the outer-core, we think of these protons as forming a solid type II superconducting surface, that forms an interface between the liquid outer-core and the inner-crust of the pulsar. We call this region the outer-core surface and guess its thickness to be less than $1\ km$, see Figure 2. The lower critical field and upper critical field of the outer-core surface is given by (2) and (3) respectively.

\begin{figure}[ht!]
\includegraphics[width=0.9\textwidth]
 {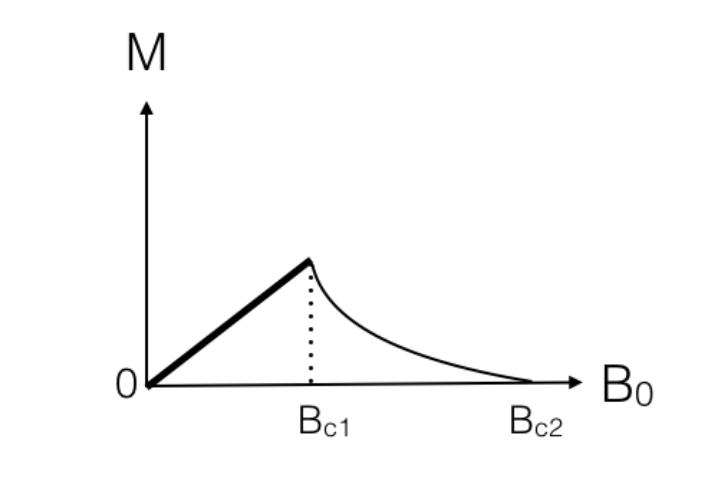}
\centering
\vspace*{-8mm}\caption{For Type II, the magnetization decreases gradually between the lower critical field $B_{c1}$ and the upper critical field $B_{c2}$, allowing magnetic flux lines to penetrate. These flux lines are called magnetic vortices.}
  \end{figure}

\begin{figure}[ht!]
\includegraphics[width=1.1\textwidth]
  {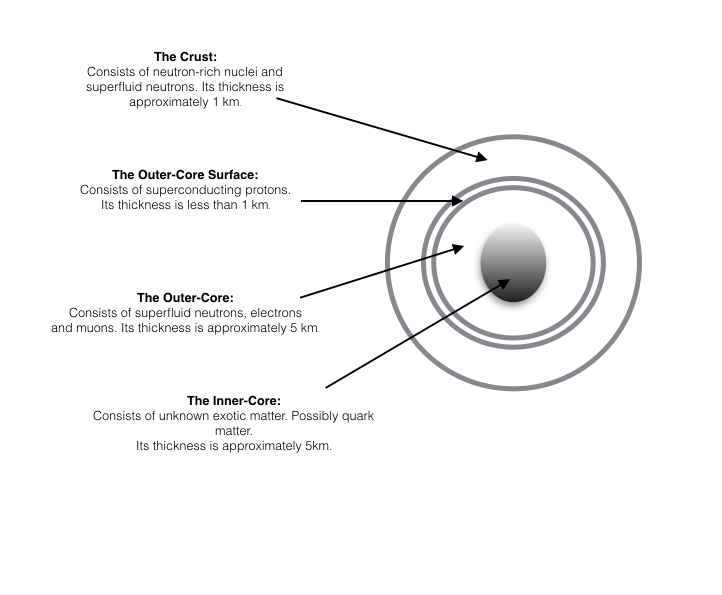} 
  \centering
  \vspace*{-29mm}\caption{This is our proposed pulsar cross-section. The currently accepted cross-section takes a similar form accept it doesn't have the outer-core surface. }
 \end{figure}

\section{AS THE PULSAR SPINS, MAGNETIC VORTICES PENETRATE THE OUTER-CORE SURFACE}

\noindent \begin{itemize}
    \item The rotation of the neutron star will cause a triangular array of vortices in the neutron superfluid. These form parrallel to the axes of rotation and have an areal number density given by \cite{charbonneau2007novel}, \begin{equation}
        n_{v}= \frac{2 \ \Omega}{\Gamma_{0}} = 10^{10} \ m^{-2},
    \end{equation} where $\Omega$ is the angular velocity of the neutron star, $\Gamma_{0} =  \pi \hbar / m_{p}$ and $m_{p} \approx 1.672621 \times 10^{-27} \ kg$ is the mass of a superconducting proton. The average neutron vortex line spacing is \cite{charbonneau2007novel}, \begin{equation}
        l_{v} = n_{v}^{-1/2} = 10^{-5}\ m.
    \end{equation}  
   
   \item The magnetic field that penetrates the proton superconductor does so by forming magnetic flux tubes with a density given by \cite{charbonneau2007novel}, \begin{equation}
       n_{f}= \frac{B}{\phi_{0}} = 10^{16} \ m^{-2}. 
   \end{equation} 
   
   \item The spacing between the proton vortices is \cite{charbonneau2007novel},
   \begin{equation}
       l_{f}= n_{f}^{-1/2} = 10^{-8} \ m.
   \end{equation}
    
    Therefore there is a higher density of proton vortices than neutron vortices. Neutron vortices become entangled in the proton vortices so this means they could only move as a single object.
    
    \item The radius of a proton vortex is given by the proton coherence length, \begin{equation} \xi_{p} \approx 30 \times 10^{-15} \ m. \end{equation}
    The London penetration depth for a proton superconductor is \cite{charbonneau2007novel}, \begin{equation} \lambda_{p} \approx 80 \times 10^{-15} \ m. \end{equation} Therefore the Ginzburg-Landau parameter for nuclear matter in a pulsar is, \begin{equation}
        \kappa = \frac{\lambda_{p}}{\xi_{p}} \approx 2.7.
    \end{equation} This of course indicates that the protons form a type II superconductor and the flux tubes have a complicated twisted structure and that the superfluid neutron vortices have many proton vortices tangled around them.

\end{itemize}

\section{AS THE PULSAR SPINS DOWN, ITS SURFACE MAGNETIC FIELD INCREASES WHICH CORRESPONDS TO AN INCREASE IN THE NUMBER OF MAGNETIC VORTICES THAT PENETRATE THE OUTER-CORE SURFACE. THIS PROCESS CONTINUES UNTIL THE UPPER CRITICAL FIELD IS REACHED, AT WHICH POINT THE OUTER-CORE SURFACE BECOMES NORMAL CONDUCTING AND THE ENERGY OF THE MAGNETIC VORTICES IS TRANSFERRED TO THE OUTER-CORE SURFACE, WHICH CAUSES A PULSAR GLITCH. WE ALSO POSTULATE THAT THIS PROCESS IS RELATED TO THE BRADLOW BOUND OF THE OUTER-CORE SURFACE  }

\newpage
\begin{itemize}
    \item The basic observed quantities of a pulsar are the angular velocity $\Omega $, the spin period $P$ and the period derivative $\dot{P}$. The formula for the surface magnetic field of a pulsar is given by \cite{konar1997magnetic, wasserman2003precession, akgun2008toroidal},
    \begin{equation} 
    B_{s} = 3.2 \times 10^{19} \bigg( P \ \dot{P} \bigg)^{1/2} \ G,
        \end{equation} where $P = 2 \pi / \Omega$, which shows that during the spin down of a pulsar $P$ increases. Therefore if we assume that $\dot{P}$ is constant we can see from (13) that when $P$ increases $B_{s}$ increases as well. 
    
    \item If we let $B_{s} = B$ in equation (8) we get,
    \begin{equation}
        n_{f} = \frac{B_{s} q}{h c},
    \end{equation} which shows that when the pulsar spins down the density of proton vortices increases. Therefore we specifically interpret this to mean that, as the pulsar spins down the number of magnetic vortices that penetrate the outer-core surface increases. Because the outer-core surface is a type II superconductor, this process will continue until the upper critical field of the outer-core surface is reached, at which point the outer-core surface will become normal conducting and the energy of the magnetic vortices will be transferred to the outer-core surface.  
    
    \item The approximate Bradlow bound for the outer-core surface is given by the formula, \begin{equation}
        N \leq \frac{A_{ns}}{A_{f}} 
    \end{equation} where $A_{ns} = 4 \pi R^{2}$ is the area of the outer-core surface and $A_{f} = \pi \xi_{p}^{2}$ is the area of a single magnetic vortex. Using $R \approx 11 000\ m$ and (10) we get that (15) becomes, \begin{equation}
      N \leq 5.3777777 \times 10^{35}.
    \end{equation}
  \item The energy of a single magnetic vortex is given by \cite{charbonneau2007novel}, \begin{equation}
      E \approx \frac{\pi \hbar^{2}}{2m_{p}} \frac{\mu}{a} \log \bigg( \frac{\Lambda}{\xi_{p}} \bigg) ,
  \end{equation} where, 
  \begin{equation} \frac{\mu}{a} \approx 8 \times 10^{43} \ m^{-3}, \end{equation} 
  is the density of superconducting protons and $\Lambda$ is a cut-off value in the integral to obtain the energy (17). We let, \begin{equation} \Lambda \approx (\xi_{p} + l_{f}) = 1 \times 10^{-8} \ m. \end{equation}
  
  \item When we substitute the corresponding values into (17) we get, \begin{equation}
        E \approx 4614.547377 \ J.
    \end{equation} When we multiply the Bradlow bound (16) with the energy of a single magnetic vortex (20) we get an approximate value of the energy $\Delta E$ that is transferred to the outer-core surface when the upper critical field is reached, \begin{equation} \Delta E \approx 2.481600998 \times 10^{46} \ ergs.
        \end{equation} The energy required to drive a pulsar glitch is approximately $10^{43} \ ergs$ \cite{larson2002simulations, tong2016rotational, lander2013contrasting}, and the energy $\Delta E$ in (21) is larger than this value, which gives support to the mechanism we present here.

\end{itemize}
\newpage
\section{CONCLUSION}

\noindent We introduced a new mechanism for pulsar glitching. We presented our mechanism in the form of three assertions/statements given as headings of section 2, 3 and 4. We gave convincing arguments in support of each statement. In closing, we propose that an experiment be conducted to test our mechanism. Even though the physical conditions in the interior of a neutron star are extreme and very different from terrestrial conditions, we still think this mechanism can be tested on earth. Even though we're not exactly sure what the conditions must be like on earth for this experiment to be successful, we do suggest the following: An applied, spherically symmetric magnetic field be placed at the center of a type II superconducting spherical shell of arbitrary thickness and size. The spherical shell must have no holes in it. The applied magnetic field must then be adjusted by remote. When the applied field reaches the upper critical field of the shell
we expect the shell to spin up. According to our mechanism, the applied field reaching the upper critical field of the shell is equivalent to the number of magnetic vortices-that penetrate the surface of the shell-reaching the Bradlow bound of the shell.

\section*{ACKNOWLEDGEMENTS}

 \noindent I thank Jesus Christ my Lord and Savior for everything. I also thank The Department of Mathematics and Applied Mathematics at the University of Cape Town.

\newpage
\section*{References}

\bibliography{mybibfile}

\end{document}